\documentclass[preprint,12pt]{elsarticle}

\usepackage{graphicx}
\usepackage{amssymb}
\usepackage{amsthm}
\usepackage{amsmath}

\usepackage{lineno}
\usepackage[utf8]{inputenc}
\usepackage{xcolor}
\usepackage{url}

\def\0{\mbox{\bf{0}}}

\journal{}

\begin{document}

\begin{frontmatter}


\title{Censored Data Forecasting: Applying Tobit Exponential Smoothing with Time Aggregation}


\author[label1]{Diego J. Pedregal}
\ead{diego.pedregal@uclm.es}
\address[label1]{ETSI Industrial de Ciudad Real.}
\author[label2]{Juan R. Trapero}
\address[label2]{Facultad de Ciencias y Tecnologías Químicas.}
\address{Universidad de Castilla-La Mancha and Institute of Applied Mathematics in Science and Engineering (IMACI). 13071 Ciudad Real, Spain}
\cortext[cor1]{Corresponding author. Tel.:+34 926 295 430}

\begin{abstract}
This study introduces a novel approach to forecasting by Tobit Exponential Smoothing with time aggregation constraints. This model, a particular case of the Tobit Innovations State Space system, handles censored observed time series effectively, such as sales data, with known and potentially variable censoring levels over time. The paper provides a comprehensive analysis of the model structure, including its representation in system equations and the optimal recursive estimation of states. It also explores the benefits of time aggregation in state space systems, particularly for inventory management and demand forecasting. Through a series of case studies, the paper demonstrates the effectiveness of the model across various scenarios, including hourly and daily censoring levels. The results highlight the model’s ability to produce accurate forecasts and confidence bands comparable to those from uncensored models, even under severe censoring conditions. The study further discusses the implications for inventory policy, emphasizing the importance of avoiding spiral-down effects in demand estimation. The paper concludes by showcasing the superiority of the proposed model over standard methods, particularly in reducing lost sales and excess stock, thereby optimizing inventory costs. This research contributes to the field of forecasting by offering a robust model that effectively addresses the challenges of censored data and time aggregation.
\end{abstract}
\end{frontmatter}

\graphicspath{ {plots/} }
\section{Introduction}
Demand forecasting is of paramount importance for supply chain management \cite{PETROPOULOS2022705}. Supply chain demand forecasts are provided to optimize decision making. Usually, such decisions are aligned to different forecasting horizons \cite{nahmias2015production}. In this sense, short term demand forecasts are used to optimize inventory  management \cite{nahmias2015production} by achieving a compromise between customer service level and inventory investment. This kind of forecasts have particular differences in comparison with other forecasting contexts. For example, lead time demand forecasts are required, where these forecasts are computed as the cumulative demand over such a lead time. Furthermore, the inventory system to be optimized have an important impact on the forecasting model selection process. Particularly, in case stockouts happen, the inventory system may work under a backordering or lost-sales schedule \cite{silver2017inventory}. In this work, we will focused on lost-sales situations, which imply that unmet demand is lost and, more importantly, most of the time is also unknown. Therefore, sales time series, which are usually used as a demand estimate to feed forecasting support systems, underestimate the underlying demand providing biased down demand forecasts. In other words, sales represent a constrained demand with regards to the underlying (unconstrained) demand \cite{Vandeput2020}. In other statistical contexts this problem is also known as censored data estimation \cite{Nahmias1994}.

Recently, authors in \cite{Trapero2024} proposed the Tobit Kalman Filter  (TKF) \cite{Allik2014} to face the problem of demand forecasting under lost-sales systems. This TKF in conjunction with the forecasting models expressed in a State Space framework \cite{harvey1989} is capable of handling trends, seasonality, exogenous variables, etc. One potential limitation of this approach is that, most of business softwares do not incorporate Kalman Filter in contrast to exponential smoothing forecasting models \cite{hyndmanbook2008}, which can be seen as a single source of error version of the Kalman Filter approach (multiple source of error), \cite{harvey1989}. In fact, exponential smoothing forecasting models have gained popularity among academics and practitioners alike, thanks to free libraries available in specialised software as R \cite{Svetunkov2022}, \cite{hyndman2018forecasting} among other advantages such as robustness, easy to automate, and low computational burden. Therefore, to take advantage of the exponential smoothing benefits and models taxonomy provided by \cite{hyndmanbook2008}, authors in \cite{pedregal2024tobitexponentialsmoothingenhanced} developed the Tobit ExponenTial Smoothing, where all the previous advances made for exponential smoothing could be used to deal with censored data.

The aforementioned stream of research estimated the unobserved part of the demand statistical distribution assuming that only sales data is available. However, as explained in reference \cite{Trapero2024}, an alternative research stream uses other sources of information besides sales data, when available. Several of those works (\cite{Lau1996}, \cite{Sachs2014}) used observations at different frequencies, i.e., they used hourly and daily sales data to estimate the censored demand. Essentially, this kind of approach explores the hourly pattern when there is no stockouts to extrapolate the rest of hourly demand when a stockout happens. This hourly forecast is aggreagated to form a daily forecast. In other words, we assume that the hourly demand pattern remains constant and, thus, it can be used to estimate the rest of demand hours after a stockout happens at a certain period of time. Nonetheless, the methodology proposed in \cite{Lau1996} was rather ad-hoc without a solid theoretical foundation. 

In this work, we extend the Tobit ETS model in \cite{pedregal2024tobitexponentialsmoothingenhanced} to handle the different data frequencies by using temporal hierarchies \cite{ATHANASOPOULOS2024430}, \cite{KOURENTZES2014}, \cite{VILLEGAS201829} to forecast censored data as happens in companies subject to lost-sales situations. Note that temporal hierarchies are based on aggregating time series of different frequencies, for example, quarterly time series can be aggregated to obtain annual time series. Therefore, we formulate the problem, previously studied by \citet{Lau1996} in an ad-hoc fashion, by using temporal hierarchies defined in a State Space framework which has more solid theoretical foundations and, thus, can provide maximum likelihood estimations, probabilistic forecasts, etc. Although, we will use hourly and daily data, other data frequency can be easily allocated.

The structure of this article is as follows: Section 2 formulates the problem and develops the Tobit Exponential Smoothing equations integrating temporal hierarchies. Section 3 explores different case studies ranging from a saturation problem in solar energy forecasting to inventory problems with constrained demand. Finally, Section 4 is devoted to draw the main conclusions.

\section{Tobit Exponential Smoothing with time aggregation constraints}
\subsection{Tobit Exponential Smoothing}
\label{sec:TETS}

The Tobit Exponential Smoothing model (TETS) with censoring restrictions is a particular case of a Tobit Innovations State Space system with the following general formulation:

\begin{equation}
\label{eq:sys}
\begin{array}{l}
x_{t} = Fx_{t-1}+ g \epsilon_{t}^*\\
y_t^* = w x_{t-1}+ \epsilon_t^*\\
y_t= \left\{
    \begin{array}{ll}
       y_{t}^*,  & y_{t}^* \leq Y_{max,t} \\
       Y_{max,t} & y_{t}^* > Y_{max,t}
    \end{array}
    \right.
\end{array}
\end{equation}

\noindent where $y_t$ is the censored observed output time series (sales in latter examples); $y_t^*$ is its uncensored counterpart (true demand); $Y_{max,t}$ is a known censoring level, possibly time varying (inventory level in later examples); $\epsilon_t^*$ is a scalar perturbation such that $\epsilon_t^* \sim N(0,\sigma^2)$; $x_t$ is a state vector and $F$, $w$ and $g$ are system matrices of appropriate dimensions ($g$ is the so called Kalman gain). 

Mind that (\ref{eq:sys}) is more general than just TETS because it may be any class of models that may be represented within an Innovations State Space system, like is the case of ARIMA models. Other interesting features, such as exogenous regression terms, even with time-varying parameters or heteroscedastic noise may be also treated by slight modifications of system (\ref{eq:sys}).

The Tobit Simple Exponential Smoothing representation in terms of system (\ref{eq:sys}) is as easy as considering the state vector as the scalar level ($x_t=l_{t}$) and taking $w=1$, $F=1$ and $g=\alpha$, where $\alpha$ is the smoothing constant such that values closed to zero produce smoother solutions.

A Tobit Holt-Winters model is a bit more complicated, though still manageable. Equation (\ref{eq:hw}) unfolds the full set of equations of such a model with a seasonal period of $m$ observations per year. The states $l_t$, $b_t$ and $s_t$ stand for the level, slope and seasonal components, respectively.

\begin{equation}
\label{eq:hw}
\centering
\begin{array}{rl}
l_{t} &= l_{t-1}+b_{t-1}+ \alpha \epsilon_{t}^* \\
b_{t} &= b_{t-1}+ \beta \epsilon_{t}^* \\
s_{t} &= s_{t-m}+ \gamma \epsilon_{t}^* \\
y_t &= l_{t-1}+b_{t-1}+s_{t-m}+ \epsilon_t^*, \quad \quad \epsilon_t \sim N(0,H) \\
\end{array}
\end{equation}

This model may be fit into a TETS in equation (\ref{eq:sys}) by setting the system matrices as indicated in equation (\ref{eq:hwSS}), that are actually the same than for the standard ETS version of the Holt-Winters model. Vertical and horizontal lines in matrices indicate the separation between the terms that define the level state (left hand) and the seasonal component (right). The unknown parameters are the smoothing constants ($\alpha$,$\beta$) and the noise variance ($\sigma^2$).

\begin{equation}
\label{eq:hwSS}
\centering
\begin{array}{rl}
x_t= & \left[ \begin{array}{cc|cccc} l_t & b_t & s_t & s_{t-1} & \dots & s_{t-m+1} \end{array} \right]^T  \\
w= & \left[ \begin{array}{cc|cccc} 1 & 1 & 0 & \dots & 0 & 1 \end{array} \right] \\
g= & \left[ \begin{array}{cc|cccc} \alpha & \beta & \gamma & 0 & \dots & 1 \end{array} \right]^T \\
F= & \left[ \begin{array}{cc|ccccc}
1 & 1 & 0 & 0 & \dots & 0 & 0 \\
0 & 1 & 0 & 0 & \dots & 0 & 0 \\
\hline
0 & 0 & 0 & 0 & \dots & 0 & 1 \\
0 & 0 & 1 & 0 & \dots & 0 & 0 \\
0 & 0 & 0 & 1 & \dots & 0 & 0 \\
0 & 0 & 0 & 0 & \dots & 0 & 0 \\
0 & 0 & 0 & 0 & \dots & 1 & 0
\end{array} \right] \\
\end{array}
\end{equation}

The optimal recursive estimation of states for the general model in equation (\ref{eq:sys}) and the estimation of the unknown parameters are developed in \cite{pedregal2024tobitexponentialsmoothingenhanced}.

\subsection{Time aggregation in Tobit Innovations State Space systems}
\label{sec:aggTETS}

One main advantage of the state space set up is that, once a model is expressed as a particular case of (\ref{eq:sys}), time aggregation can be handled as described in what follows (see e.g., \citep{harvey1989} for general state space systems).

Assume the aggregation period is $s$, e.g., in later examples hourly observations will be aggregated to daily data with $s=12$ (hours of business every day). One first step consists of writing system (\ref{eq:sys}) adding the observation equation as an additional line in the state equation, see (\ref{eq:sysone}) and assume $C_t=0$. In that case the whole system (\ref{eq:sys}) is written exactly in the form of a transition equation. There $y_{A,t}^*$ stands for a cumulative time aggregated version of $y_t^*$. If $C_t=0$ then $y_{A,t}^*=y_t^*$.

\begin{equation}
\label{eq:sysone}
    \left[ \begin{array}{c} x_t \\ \hline y_{A,t}^* \end{array} \right] =
    \left[ \begin{array}{c|c} F & 0 \\ \hline w & C_t  \end{array} \right]
    \left[ \begin{array}{c} x_{t-1} \\ \hline y_{A,t-1}^* \end{array} \right] +
    \left[ \begin{array}{c} g \\ \hline 1 \end{array} \right]
    \epsilon_t^*
\end{equation}

The inclusion of time aggregation is precisely by defining the cumulator variable $C_t$ appropriately. Take the following definition:

\[
C_t= \biggl\{ \begin{array}{ll}
        0, & t=js+1, \quad j=0,1,2,\dots \\
        1, & \text{otherwise.}
    \end{array}
\]

In that case, the the state $y_{A,t}^*$ is an accumulated version of $y_t^*$ during a cycle of $s$ observations and is reset to $y_t^*$ every time a new cycle starts over again.

The system is completed with the observation equation in system (\ref{eq:sys}). For the sake of clarity, the full TETS with time aggregation is stated as system (\ref{eq:sysfull}), where $y_{A,t}$ is the observed accumulated censored variable (sales later on) within any cycle and $Y_{max,t}$ is the censoring level for the whole cycle that applies, obviously, to the accumulated uncensored variable (true demand) $y_{A,t}^*$.

\begin{equation}
\label{eq:sysfull}
\setlength{\jot}{10pt}
\begin{array}{l}
    \left[ \begin{array}{c} x_t \\ \hline y_{A,t}^* \end{array} \right] =
    \left[ \begin{array}{c|c} F & 0 \\ \hline w & C_t  \end{array} \right]
    \left[ \begin{array}{c} x_{t-1} \\ \hline y_{A,t-1}^* \end{array} \right] +
    \left[ \begin{array}{c} g \\ \hline 1 \end{array} \right]
    \epsilon_t^*
\\ \noalign{\vskip5pt}
    y_t^* = w x_{t-1}+ \epsilon_t^*
\\ \noalign{\vskip5pt}
    y_{A,t}= \left\{
    \begin{array}{ll}
       y_{A,t}^*,  & y_{A,t}^* \leq Y_{max,t} \\
       Y_{max,t} & y_{A,t}^* > Y_{max,t}
    \end{array}
    \right.
\\ \noalign{\vskip5pt}
    C_t= \biggl\{ \begin{array}{ll}
        0, & js+1, \quad j=1,2,\dots \\
        1, & \text{otherwise.}
    \end{array}
\end{array}
\end{equation}

This system offers an elegant alternative for bottom-up time estimation and forecasting of hierarchical systems \citep{hierarchical2017} with censored data. One interesting feature of equations \citep{hierarchical2017} is that the censorship restriction is defined at a different frequency than the sampling frequency. Additionally, similar approaches can be applied to top-down or reconciled methods \citep[see e.g.,][]{VILLEGAS2018}.

\subsection{Epitome}

The problem outlined in subsection~\ref{sec:TETS} can be understood as the estimation of a signal exceeding a saturation level, beyond which the signal becomes undetectable. A typical example is presented in Figure~\ref{fig:simulation} below. For full details, refer to subsection~\ref{sec:case1}.

In contrast, subsection~\ref{sec:aggTETS} addresses a censored demand problem within a typical newsvendor setting with lost sales. Replenishment occurs daily, but sales (or demand) information is available hourly. Consequently, if a stockout happens during the day, remaining sales for that day are lost and consequently are zero. Therefore, although data is collected hourly, the inventory level for the next day should be based on the aggregated forecast for the entire day, necessitating time aggregation. See an example in Figure~\ref{fig:dailyCensoring} and details in subsection~\ref{sec:case2}.

Crucially, the solution for the newsvendor censored demand problem (i.e., system (\ref{eq:sysfull})) can be expressed as the solution to the saturation problem (i.e., system (\ref{eq:sys})), provided we choose appropriate system matrices. Taking the definitions in (\ref{eq:definitions}), system (\ref{eq:sysfull}) may be rewritten as in equation (\ref{eq:sysAll}). 

\begin{equation}
\label{eq:definitions}
\begin{array}{llll}
    x_t^+ = \left[ \begin{array}{c} x_t \\ \hline y_{A,t}^* \end{array} \right] &
    F_t^+=
    \left[ \begin{array}{c|c} F & 0 \\ \hline w & C_t  \end{array} \right] &
    g^+=
    \left[ \begin{array}{c} g \\ \hline 1 \end{array} \right] &
    w_t^+ = \left[ \begin{array}{c|c} w & C_t \end{array} \right]
\end{array}
\end{equation}

\begin{equation}
\label{eq:sysAll}
\begin{array}{rl}
x_{t}^+ &= F_{t}^+ x_{t-1}^+ + g^+ \epsilon_{t},  \\
y_{A,t} &= w_t^+ x_{t-1}^+ +\epsilon_{t}, \quad \quad \epsilon_{t} \sim N(0,H) \\
y_{A,t}&= \left\{
    \begin{array}{ll}
       y_{A,t}^*,  & y_{A,t}^* \leq Y_{max,t} \\
       Y_{max,t} & y_{A,t}^* > Y_{max,t}
    \end{array}
    \right.
\end{array}
\end{equation}

Such system is like (\ref{eq:sys}) with two small differences, namely, the transition matrices $F_t^+$ and $w_t^+$ are now time-varying, and the constraints now are operating on the estimated aggregate demand and sales. 

In essence, despite their conceptual differences, the methods to solve these two problems are fundamentally the same. The same algorithmic solutions applied to system (\ref{eq:sys}) can be used to solve system (\ref{eq:sysAll}), see the detailed algorithm in \cite{pedregal2024tobitexponentialsmoothingenhanced}.

\section{Case studies}

\subsection{Case 1: hourly saturation}
\label{sec:case1}

The first case analysed consists of a signal simulated according to an ETS(ANA) model, following the nomenclature used by \citet{hyndmanbook2008}, with Additive Error, No Trend and Additive Seasonality. In this case, we assume a saturation effect. This kind of effect is common in engineering equipment. One example is in solar power forecasting, where inverters cannot handle oversized field of panels (\cite{en17133240}).

\begin{figure}[htbp]
\centering
    \includegraphics[width=1\textwidth]{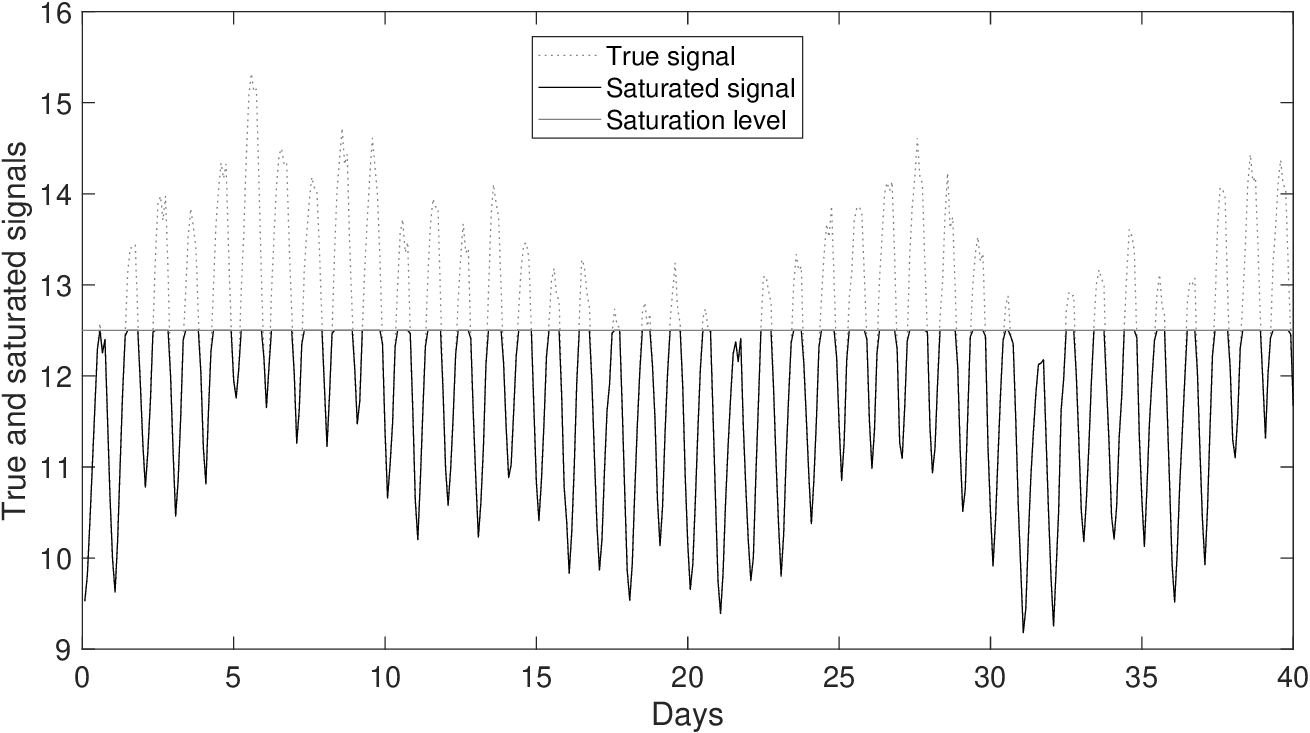}
    \caption{Simulation of true signal (grey) according to an ETS(ANA) model. Saturated signal in black for a saturation level of 12.5.}
    \label{fig:simulation}
\end{figure}

In this type of application, there is no time aggregation, since we are dealing only with the saturation case outlined in subsection \ref{sec:TETS}. The model includes an additive noise, a local level trend and additive seasonality (see e.g., \cite{hyndmanbook2008}), with simulated parameters $\alpha=0.99$, $\gamma=0.006$ and $\sigma^2=0.02$. Figure \ref{fig:simulation} shows the true signal in dotted-grey colour, saturated signal in black assuming a saturation level of 12.5. Mind that signal values above 12.5 are never observed. Note that this saturation level is rather restrictive because some parts of the signal (between days 5-10) are almost lost. 

\begin{figure}[htbp]
\centering
    \includegraphics[width=1\textwidth]{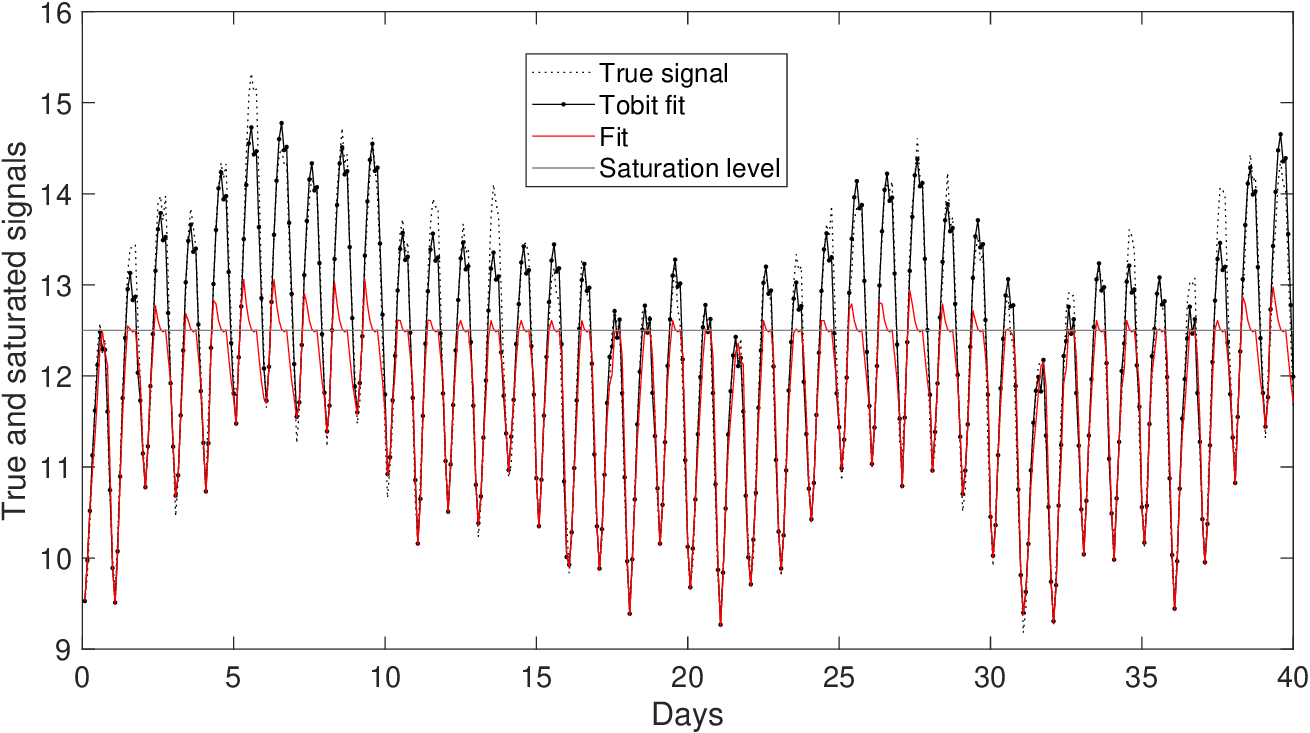}
    \caption{True signal (dotted), standard fit using saturated signal (red solid) and Tobit fit (dashed dot).}
    \label{fig:initialFit}
\end{figure}

Figure \ref{fig:initialFit} shows the standard and Tobit fits for a censoring level of 12.5 operating at every single hour. The bias when the censoring level is not taken into account is evident, since it is unable to fit values above the censoring level properly. The Tobit ETS model, estimated according to the initial system (\ref{eq:sys}), clearly fits the data above the censoring level. 

\subsection{Case 2: daily censoring level}
\label{sec:case2}

\begin{figure}[htbp]
\centering
    \includegraphics[width=1\textwidth]{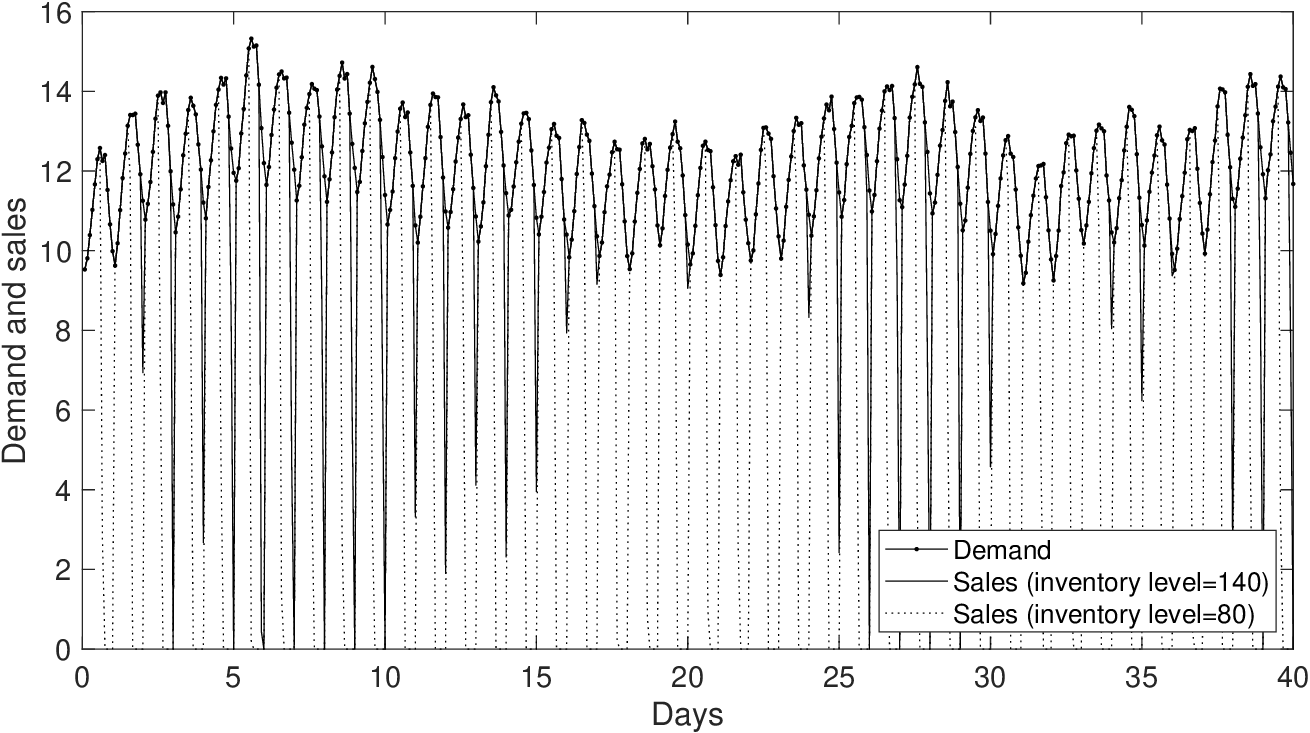}
    \caption{Demand (dot-solid) and censored sales with $Y_{max,t}=122$ (solid) and wih $Y_{max,t}=80$ (dots).}
    \label{fig:dailyCensoring}
\end{figure}

A more realistic case that requires time aggregation models like (\ref{eq:sysfull}) or (\ref{eq:sysAll}) is when a censoring level is operating daily with hourly sampled data, that is equivalent to assume that the stock is replenished every day but sales recorded per hour. Every stockout produced during a day is refilled at the beginning of the next day. Figure \ref{fig:dailyCensoring} shows demand and sales in such a case for two daily censoring levels, namely 140 and 80. The simulated demand uses the same ETS(ANA) time series than in the previous example. The figure shows some spikes at the end of some days when the stockouts are produced. A zero sales value indicates that the stockout happened at some time previous to the last hour, while non-zero values departing from true demand is evidence that the stockout happened at the very last hour of the day. Sales and demand lines are overlapped until stockout happens. Judging by this, the censorship is not very restrictive in this case of 140, but is much more strict in the case of 80.

It is evident that, in either case, the estimation using the bare sales time series would tend to replicate the spikes, producing rather biased estimates. This is especially important regarding the demand variance estimate, that is crucial for defining the safety stock of the inventory management \cite{Trapero2019a}, as shown in subsection \ref{sec:case3}.

\begin{figure}[htbp]
\centering
    \includegraphics[width=1\textwidth]{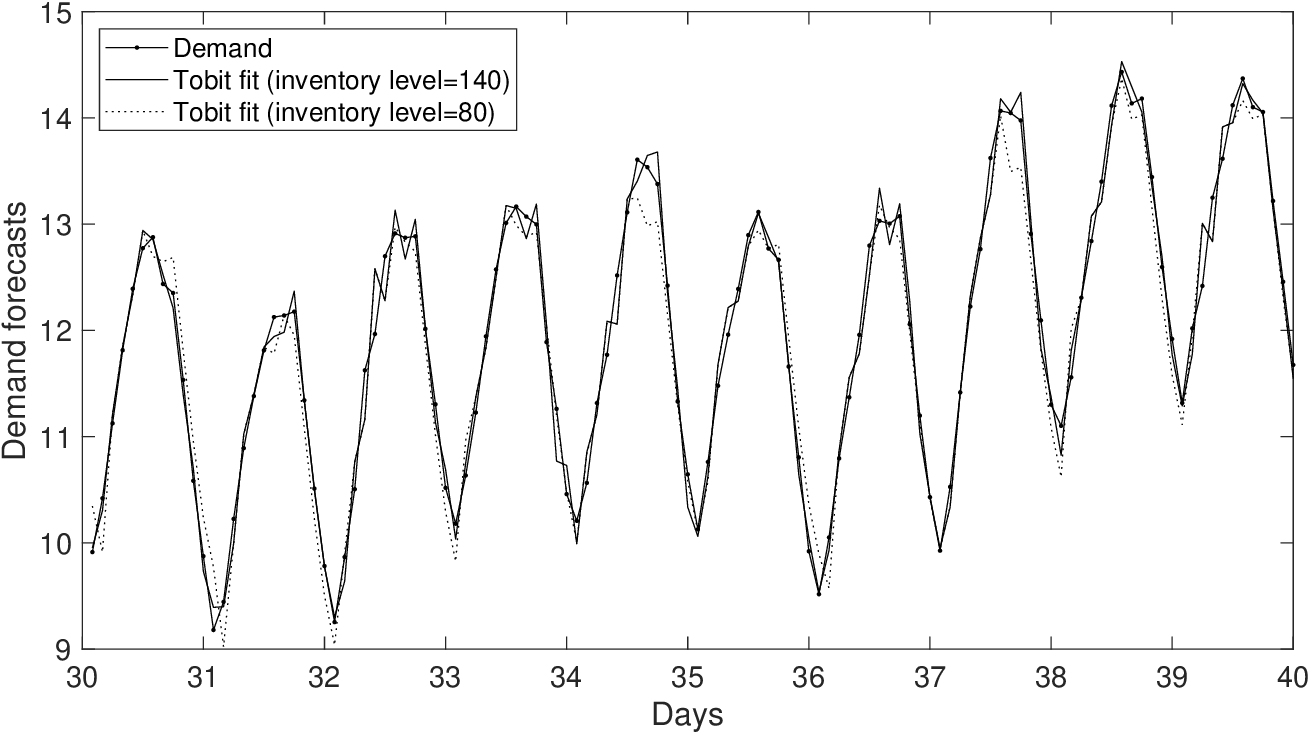}
    \caption{Demand (dot-solid), and Tobit fits with two different degrees of censorship.}
    \label{fig:dailyFit}
\end{figure}

Fitted values from a Tobit model in equation (\ref{eq:sysfull}) or (\ref{eq:sysAll}) with different levels of censorship produces the estimates shown in Figure \ref{fig:dailyFit}, where no spikes at all are devised.

\begin{figure}[htbp]
\centering
    \includegraphics[width=1\textwidth]{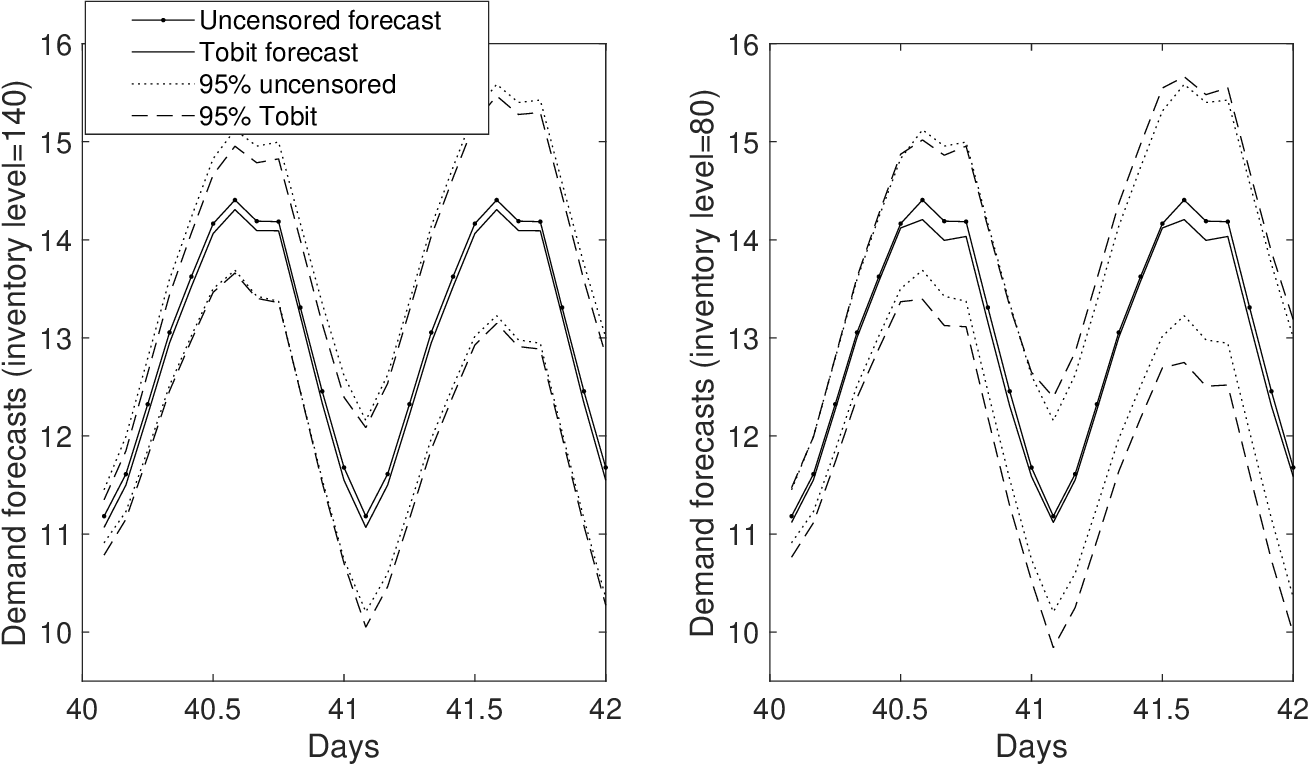}
    \caption{Forecasts and 95\% confidence bands with uncensored model on true demand (dot-solid and dotted) and the Tobit forecasts with different degrees of censoreship ($Y_{max,t}=140$ on the left and $Y_{max,t}=80$ on the right).}
    \label{fig:dailyForecasts}
\end{figure}

Most interestingly, forecasts and confidence bands produced are very similar to the ones obtained in the case that the true demand would be known and forecasted with a standard model with no time aggregation. This is shown in Figure \ref{fig:dailyForecasts}, where the forecasts with different censoring severity are shown together with the forecasts estimated when the true demand is fully known.

\subsection{Case 3: A Newsvendor inventory simulation}
\label{sec:case3}
In this case, more complex examples originated from supply chain management are simulated, where it is assumed a lead time of one day and a determined target Cycle Service Level (CSL). The lead time is the period between the time an order is placed and the time when it is fulfilled. The simulated inventory policy follows a Newsvendor system. The safety stock is designed based on a certain target CSL, where such a CSL is estimated as the corresponding percentile of the forecasting demand distribution \cite{Trapero2019a}. Then, the quantile produced by the point forecast plus the safety stock is the censoring level for the lead time (one day) and is continuously changing over time. The censoring level determines the lost sales and excess stock at each day and the sales are built according to the constraints in (\ref{eq:sysfull}) or (\ref{eq:sysAll}) and used in later forecasts.

More specifically, the simulations consist of the following steps:

\begin{enumerate}
    \item Select a target CSL.
    \item Estimate the daily point and distribution forecasts by three models: i) on daily data with a standard ETS model based on equation (\ref{eq:sys}) but without constraints (called ETS herein), ii) on daily data with a Tobit model (\ref{eq:sys}) (TETS), iii) on hourly data with a Tobit model and time constraints (\ref{eq:sysfull}) or (\ref{eq:sysAll}) (TETSC).
    \item Set the inventory level for next day ($Y_{max,t}$) at the service level percentile selected in 1. for the three models in 2. and forecast sales for next day. Sales equal demand when demand is smaller or equal to $Y_{max,t}$, but sales are truncated to $Y_{max,t}$ when demand is greater. These estimated sales are the ones used in next steps.
    \item Estimate excess inventory and lost sales with respect to the true demand for ETS, TETS and TETSC.
    \item Repeat steps 2. to 4. for a whole year of data, i.e., 365 rounds of 1 day ahead forecasts for ETS and TETS and 365 rounds of 12 hours ahead forecasts for TETSC.
\end{enumerate}

Comments:

\begin{itemize}
    \item Method i) in step 2. above would be the standard procedure for many practitioners in industry.
    \item This simulation produces different estimates of demand, lost sales and excess inventory depending on the model used. Therefore, the historical time series fed in the models in step 2. are different for each of the models.
    \item One key issue here is realising that inventory policy depends on forecasts (i.e., $Y_{max,t}$ changes daily based on forecasts) and forecasts depend on the inventory policy (since sales are estimated based on sales or censored demand).
\end{itemize}

\subsubsection{Spiral-down effects}
One difficulty that immediately appears when the censoring essence of the forecasts is ignored in the ETS model is the spiral-down effects, as is well documented in the literature, see e.g., \cite{Vandeput2020}, \cite{Cooper2006}. As a consequence of demand downward biased estimation when stockout occur, a lower inventory level and a lower customer service level are in place, meaning that the actual customer service level departs further away from its target as time passes by.

\begin{figure}[htbp]
\centering
    \includegraphics[width=1\textwidth]{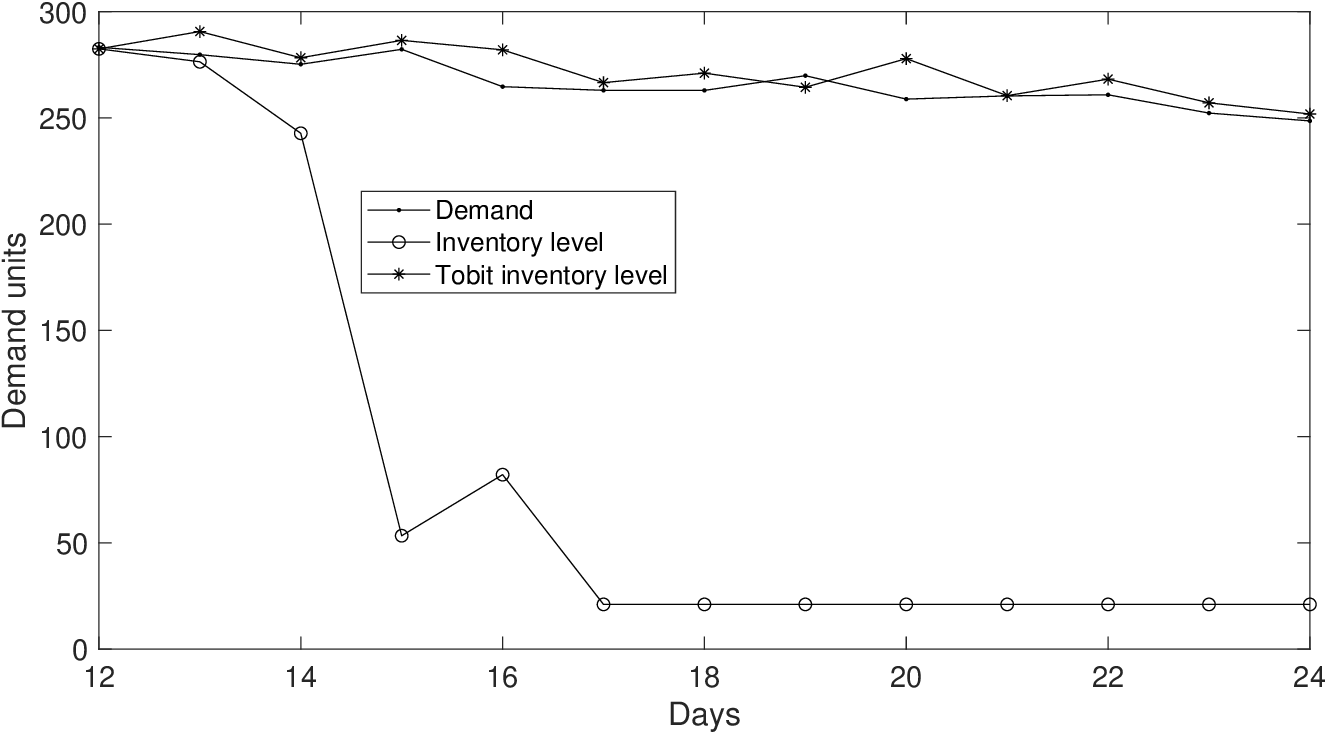}
    \caption{Stock levels for standard ETS and TETS models.}
    \label{fig:spiralDown}
\end{figure}

This effect is clearly seen in Figure \ref{fig:spiralDown} where the inventory level ($Y_{max,t}$ in the models) estimated with model ETS (circles) deteriorates systematically once a stockout happen, incurring in many accumulated lost sales. Such drastic situation is produced when models are applied blindly, and the situation is clearly not realistic, since any inventory manager would apply `judgemental' corrections to forecasts in order to avoid such a blunt policy relying blindly on forecasting models. The definite loser here are the forecasting model and forecasters, but the point is that the model is terribly wrong.

\subsubsection{Forecasting performance}
How better Tobit is with respect to the corrected version is shown in tables \ref{tab:forecastingMetrics} and \ref{tab:lostSalesetc}, where the three algorithms have been run for a full year of data (365 days). 

Table \ref{tab:forecastingMetrics} shows standard forecasting metrics, namely Root Mean Squared Error (RMSE) and the bias of forecasts, to compare the performance of the three models considered (ETS, TETS, TETSC) for different target CSL ranging from 80\% to 99\%.

The conclusions are clear, since the ranking of models regarding both precision and bias is independent of the CSL taken as a target. There are significant advantages using Tobit models and such advantages are still greater for the hourly models with time aggregation (TETSC). One interesting and expected outcome is that both precision and bias improves for the ETS model as the CSL increases. This is due to the fact that greater CSLs allow for a better approximation of sales to the true demand. Note that the precision and bias obtained by both Tobit models are constant and independent of the CSL. This is an evidence of the robustness of the Tobit models against the level of censorship.

\subsubsection{Inventory performance}
Regarding the inventory performance of the forecasting models considered, Table \ref{tab:lostSalesetc} shows lost sales, excess inventory and the achieved CSL by the three models for different target CSLs. Note that those metrics are obtained after an inventory policy, as the Newsvendor one, is used in the simulation. Several outcomes are worth mentioning:

\begin{itemize}
    \item Consistently, as the target CSL increases across all models, we see a decrease in lost sales and an increase in excess inventory.
    \item Daily TETS improves daily ETS by reducing the lost sales at the cost of higher excess inventory. As the CSL rises, the gap between these two inventory levels naturally tends to shrink. This highlights the crucial role of optimizing the target CSL to keep inventory costs in check.
    \item What is remarkably important is that the hourly TETSC improves upon daily TETS (both Tobit models) by decreasing both lost sales and excess inventory, for all CSL except 99\%, though overall improvements in costs are clear even for that case.
    \item Hourly TETSC consistently reduces both lost sales and excess inventory compared to daily ETS for all target CSLs except 80\%. 
    \item Tobit models consistently deliver CSLs that are significantly closer to target CSLs. While there are minor discrepancies between TETS and TETSC methods, both achieve similar results.
\end{itemize}

This case study suggests that leveraging all available data (hourly) for forecasting and the information about stockouts could potentially enhance forecasting accuracy and thus, inventory performance at other granularity levels (e.g., daily).

      
\begin{table}[h]
    \centering
    \begin{tabular}{lrrrr}
    \hline
  &  80\%  & 90\%  & 95\% & 99\% \\
\hline
RMSE ETS &  25.0 & 12.2 & 10.7 & 10.2 \\
RMSE TETS & 10.9 & 10.2 & 10.1 & 10.1 \\
RMSE TETSC & 3.9 & 4.0 & 3.9 & 3.8 \\
\hline
Bias ETS &  -12.0 & -3.5 & -2.1 & -1.6 \\
Bias TETS &  -1.8 & -1.5 & -1.4 & -1.4 \\
Bias TETSC & -0.3 & -0.2 & -0.2 & -0.3 \\
    \hline
    \end{tabular}
    \caption{Performance metrics for the case study IV for different target cycle service levels.}
    \label{tab:forecastingMetrics}
\end{table}
                   
 \begin{table}[h]
    \centering
    \begin{tabular}{lrrrr}
    \hline
  &  80\%  & 90\%  & 95\% & 99\% \\
\hline
Lost sales ETS &  3,522 & 690 & 229 & 21 \\
Lost sales TETS & 639 & 261 & 126 & 12 \\
Lost sales TETSC & 393 & 186 & 94 & 21 \\
\hline
Excess inventory ETS & 1,596 & 3,289 & 4,678 & 7,226 \\
Excess inventory TETS & 2,914 & 4,178 & 5,191 & 7,404 \\
Excess inventory TETSC & 2,322 & 3,294 & 4,094 & 5,733 \\
\hline
Achieved CSL ETS & 50.9\%  & 76.0\% & 87.7\% & 97.4\% \\
Achieved CSL TETS & 75.1\%  & 85.4\% & 93.1\% & 98.6\% \\
Achieved CSL TETSC & 72.8\%  & 87.1\% & 93.0\% & 98.0\% \\
    \hline
    \end{tabular}
    \caption{Lost sales, excess inventory and achieved CSL for different values of target CSL.}
    \label{tab:lostSalesetc}
\end{table}

\section{Conclusions}
\label{sec:conclusion}
Demand forecasting is a crucial task for optimizing the supply chain. Typically, sales is recorded to be used as a demand estimate that is fed into forecasting support system. However, in the presence of stockouts sales underestimate demand and, thus, a statistical method should be employed to correct that bias. Despite of the practical interest, very few ad-hoc methods have been explored until recently, where different State Space models that considered censored data have been developed. In this work, a novel hierarchical State Space model has been developed to deal with censored data at different frequencies. This new model has been shown to be useful for supply chain applications where observations are stores at a higher frequency than the censorship applies. The results show that the forecasting performance of the proposed model is remarkable superior. 

In the case study analyzed, a newsvendor inventory policy was simulated. The promising results obtained in forecasting terms were also translated into inventory terms reaching a decrease of lost sales even with a lower inventory investment. In addition, the importance of correctly handling censored demand data was evident with the analysis of spiral-down effects. 

Further research should extend these forecasting model for more complex inventory policies. In addition, empirical case studies using real data from different industries are also required to corroborate the results shown in this work.


\section{Acknowledgements}
\label{acknow}
This work was supported by the Vicerrectorado de Investigación y Política Científica from UCLM through the research group fund program (PREDILAB; DOCM 2023-4428 [2022-GRIN-34368]).




\bibliographystyle{model1-num-names}
\bibliography{biblio.bib}







\end{document}